\documentclass[final,authoryear,5p,times,twocolumn]{elsarticle} 

\usepackage{graphicx}
\usepackage{amssymb}
\usepackage{color}
\usepackage{url}
\usepackage{multirow}
\usepackage{endfloat}

\journal{Economics Letters}

\begin{document}

\begin{frontmatter}

\title{Unveiling correlations between financial variables and topological metrics of trading networks: Evidence from a stock and its warrant}
\author[BS,RCE,SS]{Ming-Xia Li}
\author[BS,RCE]{Zhi-Qiang Jiang}
\author[BS,RCE,SS]{Wen-Jie Xie}
\author[TJU,CCSCA]{Xiong Xiong}
\author[TJU,CCSCA]{Wei Zhang\corref{cor1}}
\ead{weiz@tju.edu.cn}%
\author[BS,RCE,SS]{Wei-Xing Zhou\corref{cor1}}
\cortext[cor1]{Corresponding author. Address: 130 Meilong Road, P.O. Box 114, School of Business,
              East China University of Science and Technology, Shanghai 200237, China; Telephone: +86-21-64253634.}
\ead{wxzhou@ecust.edu.cn}%

\address[BS]{School of Business, East China University of Science and Technology, Shanghai 200237, China}
\address[RCE]{Research Center for Econophysics, East China University of Science and Technology, Shanghai 200237, China}
\address[SS]{Department of Mathematics, East China University of Science and Technology, Shanghai 200237, China}
\address[TJU]{College of Management and Economics, Tianjin University, Tianjin 300072, China}
\address[CCSCA]{China Center for Social Computing and Analytics, Tianjin University, Tianjin 300072, China}

\begin{abstract}
  Traders adopt different trading strategies to maximize their returns in financial markets. These trading strategies not only results in specific topological structures in trading networks, which connect the traders with the pairwise buy-sell relationships, but also have potential impacts on market dynamics. Here, we present a detailed analysis on how the market behaviors are correlated with the structures of traders in trading networks based on audit trail data for the Baosteel stock and its warrant at the transaction level from 22 August 2005 to 23 August 2006. In our investigation, we divide each trade day into 48 time windows with a length of five minutes, construct a trading network within each window, and obtain a time series of over 1,100 trading networks. We find that there are strongly simultaneous correlations between the topological metrics (including network centralization, assortative index, and average path length) of trading networks that characterize the patterns of order execution and the financial variables (including return, volatility, intertrade duration, and trading volume) for the stock and its warrant. Our analysis may shed new lights on how the microscopic interactions between elements within complex system affect the system's performance.

\end{abstract}


\end{frontmatter}

\newpage

\section{Introduction}
\label{section:Introduction}

Trading networks, in which the nodes represent the traders and the edges stand for the trading relationships, are much used to model the buy-sell interactions among traders in economic systems. Due to the availability of audit trail data at the transaction level, many effects have been put into the understanding of the topological characteristics of trading networks. \cite{Kyriakopoulos-Thurner-Puhr-Schmitz-2009-EPJB} find that the Austrian money flow trading network is disassortative and also exhibit low correlation between node degrees and transaction volume. \cite{Tseng-Li-Wang-2010-EPJB, Tseng-Lin-Lin-Wang-Li-2010-PA} also discover that the trading network in a web-based experimental prediction market is scale-free and disassortative. \cite{Jiang-Zhou-2010-PA} reveal that the daily trading networks in Shenzhen stock market exhibit patterns of power-law degree distributions and disassortative architectures and possess a power-law correlation between the average degree and the size of placed orders. \cite{Wang-Zhou-Guan-2011-PA} find that the trading networks in the Shanghai Futures Market exhibit similar features such as scale-free, small-world effect, hierarchical organization, and power-law betweenness distribution.

Trading networks provide an overall perspective to describe the detailed transactions between pairwise traders, which can be applied to find clues and propose approaches to track and detect the abnormal trades implemented by manipulators in financial markets. \cite{Kyriakopoulos-Thurner-Puhr-Schmitz-2009-EPJB} report that the random matrix analysis is able to identify accounts with financial misconduct. \cite{Tumminello-Lillo-Piilo-Mantegna-2012-NJP} identify clusters of traders with a very high degree of synchronization in trading, which probably is also related to price manipulation. By investigating the trading networks of manipulated and non-manipulated stocks in the Shanghai stock market, \cite{Sun-Cheng-Shen-Wang-2011-PA} and \cite{Sun-Shen-Cheng-Wang-2012-PLoS1} observe significant differences in the topological properties between manipulated and non-manipulated stocks. \cite{Jiang-Xie-Xiong-Zhang-Zhang-Zhou-2013-QFL} find that the abnormal trading motifs (self-loop, two-node loop, and two-node multiple arcs) in trading networks have connections with turbulent market dynamics and can be adopted to develop novel tools for the detection of trade-based price manipulations.

Trading networks are formed based on the transactions triggered by traders with different trading strategies, which provides two perspectives to understand the trading behaviors of traders. On the one hand, the successful strategies can be followed and copied by other traders and the traders with correlated strategies will lead to similarity both in returns and positions in trading networks. \cite{Cohen-Cole-Kirilenko-Patacchini-2011-SSRN} report that traders' returns are correlated with their positions occupied in the trading networks. They also find that the market shocks at the individual level can be widely transmitted and greatly amplified through the links of trading networks. On the other hand, the specific trading strategies of traders may lead to specific local structures in trading networks and also have some influences on market behaviors. \cite{Adamic-Brunetti-Harris-Kirilenko-2012-SSRN} explore the trading networks constructed from the September 2009 E-mini S\&P 500 futures contracts and found that there are contemporaneously correlated relationships between network metrics (centralization, assortative index and average path length) and financial variables (return, volatility, intertrade duration, and trading volume).

In this paper, we will follow \cite{Adamic-Brunetti-Harris-Kirilenko-2012-SSRN} to investigate correlations between financial variables and topological metrics of trading networks of a liquid Chinese stock and its warrant. The remainder of this paper is organized as follows. In Sec. \ref{Section:DataSets}, we briefly describe the data sets we adopt. Sections \ref{Section:TopologicalCharacteristic} and \ref{Section:MarketCharacteristic} define respectively the topological metrics of trading networks and the financial variables investigated in this work. The correlations between financial variables and topological metrics of trading networks are presented in Section \ref{Section:Correlations}. We summarize our findings in Sec. \ref{Section:Conclusion}.

\section{Data sets}
\label{Section:DataSets}

We employ a transaction-level database, which contains the actual trades completed on the Shanghai Stock Exchange for stock Baosteel and its warrant, to uncover the potential connection between the equities' behaviors and the topological structures of trading networks. The Shanghai securities market is a pure order-driven market and the observed trades are triggered by matching the market orders or executable limit orders with the orders sitting on the opposite side of limit order book, which is a waiting queue of limit orders sorted by price and time priority. There are totally 3,779,538 transactions for stock Baosteel and 25,483,344 transactions for its warrant during our sampling period from 22 August 2005 to 23 August 2006, covering 243 trading days. Each transaction contains the following information: a unique transaction ID, two unique trader IDs, the trading time, the trade size, the transaction price, and an indicator of buy or sell. We note that the stock (warrant) is traded according to the $T+1$ ($T+0$) rule.

\section{Trading network metrics}
\label{Section:TopologicalCharacteristic}

We divide each trading day into 48 time windows with an interval of 5 minutes. In each window, the transaction records are translated into one trading network, in which the nodes represent the traders and the links weighted by the trade sizes are connected from the sellers to the buyers. Three topological metrics (centralization, assortative index, and average path length) are adopted to quantitatively describe the structure of each trading network. Denote $n$ the number of traders in a window and $m$ the number of edges or transactions. We obtain that $n=154\pm170$ and $m=161\pm185$ for the stock and $n=990\pm1139$ and $m=1093\pm1299$ for the warrant. The large standard deviations of the four quantities shows that the structure of trading networks vary remarkably along time. Moreover, the warrant trading networks have more nodes and edges than the stock trading networks, implying that the trading activity of the warrant is higher than the stock.

The unweighted network centralization $C_k$ is defined as the difference between the out-degree centralization $C_{k}^{\rm{out}}$ and the in-degree centralization $C_{k}^{\rm{in}}$, such that $C_k = C_k^{\rm{in}} - C_k^{\rm{out}}$. Here we use the degree to measure the node centrality. The in-degree centralization can be estimated through the well-known definition of \cite{Freeman-1979-SN}:
\begin{equation}
\label{Eq:Ckin}
C_{k}^{\rm{in}} = \frac{\sum_{i=1}^n (k_{\max}^{\rm{in}} - k_i^{\rm{in}})}{(n-1)^2},
\end{equation}
where $k_i^{\rm{in}}$ is the in-degree of node $i$, $k_{\max}^{\rm{in}}$ is the maximal in-degree and $n$ is the number of nodes. The denominator $(n-1)^2$ in Eq.~(\ref{Eq:Ckin}) represents the maximum possible sum of differences in point centrality for a directed graph of $n$ nodes. Hence, $C_{k}^{\rm{in}}$ varies between 0 and 1. By replacing the in-degree with the out-degree, we can obtain the out-degree centralization,
\begin{equation}
\label{Eq:Ckout}
C_{k}^{\rm{out}} = \frac{\sum_i^n (k_{\max}^{\rm{out}} - k_i^{\rm{out}})}{(n-1)^2}.
\end{equation}
$C_{k}^{\rm{out}}$ also locates in the range of $[0, 1]$.

The measure $C_k$ falls in the interval $[-1, 1]$ \citep{Newman-2002-PRL}. If a network has a star-like core-periphery structure in which all the arrows are from the peripheral nodes to the core node such that the core node acts as a sink, the in-degree and the out-degree of the sink are respectively $k_{\max}^{\rm{in}}=n-1$ and $0$, while the in-degree and out-degree of the peripheral nodes are 0 and $k_{\max}^{\rm{out}}=1$. We have $C_k^{\rm{in}}=1$ and $C_k^{\rm{out}}=1/(n-1)^2$, which leads to $C_k\approx1$ when $n$ is large. If a network a star-like core-periphery structure in which all the arrows are from the core node to the peripheral nodes and the core node acts as a source, the in-degree and the out-degree of the source are respectively 0 and $k_{\max}^{\rm{out}}=n-1$, while the in-degree and out-degree of the peripheral nodes are $k_{\max}^{\rm{in}}=1$ and $0$. We have $C_k^{\rm{in}}=1/(n-1)^2$ and $C_k^{\rm{out}}=1$, which leads to $C_k\approx-1$ when $n$ is large. These two cases are not usual and can occur when the time window is small. The cases that $C_k=\pm1$ indicate the presence of a dominating buyer and a dominating seller.

The weighted network centralization $C_s$ is extended from the unweighted network centralization $C_k$, in which the node centrality are measured by its node strength. Similar to $C_k^{\rm{in}}$, we can write the in-strength centralization as
\begin{equation}
\label{Eq:Csin}
C_{s}^{\rm{in}} = \frac{\sum_i^n (s_{\max}^{\rm{in}} - s_i^{\rm{in}})}{(n-1)\sum_i^{n} s_i^{\rm{in}}},
\end{equation}
where $s_i^{\rm{in}}$ is the in-strength of node $i$, $s_{\max}^{\rm{in}}$ is the maximum in-strength, and $n$ is the number of nodes. The out-strength centralization can be also written as
\begin{equation}
\label{Eq:Csout}
C_{s}^{\rm{out}} = \frac{\sum_i^n (s_{\max}^{\rm{out}} - s_i^{\rm{out}})}{(n-1)\sum_i^{n} s_i^{\rm{out}}}.
\end{equation}
Obviously, both $C_{s}^{\rm{in}}$ and $C_{s}^{\rm{out}}$ will vary between 0 and 1. The weighted network centralization $C_s = C_{s}^{\rm{in}} - C_{s}^{\rm{out}}$ has the same physical meaning as the unweighted network centralization $C_k$. Instead of using the number of counterparties, the advantage of $C_s$ is to indicate the dominating buyer or seller by taking the trade size into account.

A network possesses assortative (disassortative) mixing patterns if its high-degree nodes tend to be connected to high-degree (low-degree) nodes. \cite{Newman-2002-PRL,Newman-2003-PRE} has proposed a measure to quantitatively capture the mixing patterns of directed networks. We simply adopt this measure to see how ask (respectively, bid) nodes are surrounded with bid (ask) nodes, because we also observe that one trader corresponds only to either a seller or a buyer in most of the five-minute trading networks. The assortative index is written as follows:
\begin{equation}
\label{Eq:rk}
e_k = \frac{\sum_{j=1}^m \frac{ek_j^{\rm{a}} ek_j^{\rm{b}}}{m} - \left[\sum_{j=1}^m\left(\frac{ek_j^{\rm{a}} + ek_j^{\rm{b}}}{2m}\right)\right]^2}
      {\sum_{j=1}^m \left[\frac{\left(ek_j^{\rm{a}}\right)^2 + \left(ek_j^{\rm{b}}\right)^2}{2m} \right] - \left[\sum_{j=1}^m\left(\frac{ek_j^{\rm{a}} + ek_j^{\rm{b}}}{2m}\right)\right]^2},
\end{equation}
where $ek_j^{\rm{a}}$ and $ek_j^{\rm{b}}$ are the degrees of the two endpoints $a$ and $b$ for the edge corresponding to trade $j$, and $m$ is the number of directed edges from the ask nodes to the bid nodes. As mentioned in \cite{Newman-2002-PRL}, $e_k$ is in the range of $[-1, 1]$ and $e_k<0$ indicates that the network is disassortative. We also have $e_k = -1$ for star networks with all the peripheral nodes pointing to or departing from the central nodes.

Taking the trading size into account, we generalize the weighted assortative index through replacing node degree with node strength in Eq.~(\ref{Eq:rk}):
\begin{equation}
\label{Eq:es}
e_s = \frac{\sum_{j=1}^m \frac{es_j^{\rm{a}} es_j^{\rm{b}}}{m} - \left[\sum_{j=1}^m\left(\frac{es_j^{\rm{a}} + es_j^{\rm{b}}}{2m}\right)\right]^2}
      {\sum_{j=1}^m \left[\frac{\left(es_j^{\rm{a}}\right)^2 + \left(es_j^{\rm{b}}\right)^2}{2m} \right] - \left[\sum_{j=1}^m\left(\frac{es_j^{\rm{a}} + es_j^{\rm{b}}}{2m}\right)\right]^2},
\end{equation}
where $es_j^{\rm{a}}$ and $es_j^{\rm{b}}$ are the strengths of the two endpoints $a$ and $b$ for the edge corresponding to trade $j$, and $m$ is the number of directed edges from the ask nodes to the bid nodes.

The average path length $l$ is defined as the mean value of the shortest path length between two arbitrary nodes in a trading network. We discard the direction and weights of the trading network while computing the path length. The smaller the average path length is, the more tightly the traders are connected in the trading network.

\setlength\tabcolsep{3.5pt}
\begin{table}[tp]
\centering
  \caption{\label{Tb:Statistics:NV} Summary statistics of trading network metrics.}
  \medskip
\begin{tabular}{cccccccccc}
  \hline\hline
     & $C_k$ & $C_s$ & $e_k$ & $e_s$ & $l$ \\
     \hline
 \multicolumn{6}{l}{Panel A: Stock} \\
 Mean & -0.0518 & -0.0313 & -0.5340 & -0.3596 & 4.7969 \\
 Median & -0.0498 & -0.0284 & -0.5212 & -0.3476 & 4.1943 \\
 Maximum & 0.9990 & 0.9145 & 0.0981 & 0.8361 & 29.319 \\
 Minimum & -0.9561 & -0.8735 & -1.0000 & -0.9929 & 1.3077 \\
 Std. Dev. & 0.3106 & 0.3075 & 0.1454 & 0.2179 & 2.5057 \\
 Skewness & 0.1063 & 0.0581 & -0.4336 & 0.0255 & 1.9568\\
 Kurtosis & 3.1840 & 2.8411 & 3.1698 & 3.6339 & 9.7857\\
 \hline
 \multicolumn{6}{l}{Panel B: Warrant} \\
 Mean & -0.0100 & -0.0036 & -0.3279 & -0.1910 & 6.9542 \\
 Median & -0.0090 & 0.0000 & -0.3239 & -0.1861 & 6.7217 \\
 Maximum & 1.0000 & 0.9456 & 0.0551 & 0.2369 & 18.840 \\
 Minimum & -0.5453 & -0.6942 & -1.0000 & -0.9998 & 1.9594 \\
 Std. Dev. & 0.0718 & 0.0801 & 0.0915 & 0.0794 & 1.5639 \\
 Skewness & 3.6731 & 0.9150 & -0.7452 & -1.5393 & 1.2668\\
 Kurtosis & 52.527 & 20.719 & 6.0069 & 16.052 & 6.8101 \\
  \hline\hline
  \end{tabular}
\end{table}

Table~\ref{Tb:Statistics:NV} reports the summary statistics of the unweighted network centralization $C_k$, the weighted network centralization $C_s$, the unweighted assortative index $e_k$, the weighted assortative index $e_s$, and the average path length $l$ of the trading networks for both the stock and its warrant. The Augmented Dickey-Fuller test demonstrates that all the network variables are stationary. The Jarque-Beta test reveals that all network metrics are not normally distributed. The Ljung-Box Q-test for all network metrics rejects the null hypothesis of no autocorrelation. In other words, the time series of these network metrics possess long memory. The fluctuations of the network metrics for the stock are larger than those for the warrant.

The unweighted network centralization $C_k$ and the weighted network centralization $C_s$ on average equals to $-0.05 \pm 0.31$ and $-0.03 \pm 0.31$ for stock trading networks. While for warrant trading networks, we have $C_k = -0.01 \pm 0.07$ and $C_s = 0.00 \pm 0.08$. All the four estimated values are very close to 0, which indicates that both sides of limit order book have very similar order execution patterns. The two maxima of $C_k$ for the stock and its warrant are very close to 1, implying the presence of a dominating buyer in each case. However, the two maxima of $C_s$ are relative smaller. It means that the trade sizes of the remainder transactions are larger than those in the dominating star structure with a source. In contrast, the differences of the minima from -1 are relatively larger. We find that these two measures are right-skewed, especially for the warrant. The kurtosis of $C_k$ and $C_s$ are close to 3 for the stock, indicating that the two time series do not have fat tails. On the contrary, the kurtosis for the warrant are very large and hence the two metrics exhibit remarkable fat tails.

We also observe that all the means of the assortative indexes are negative ($e_k = -0.53 \pm 0.15$ and $e_s = -0.36 \pm 0.22$ for stock trading networks and $e_k = -0.33 \pm 0.09$ and $e_s = -0.19 \pm 0.08$ for warrant trading networks), which implies that the trading networks exhibit a disassortative pattern. In addition, the degree of disassortativeness is stronger for the stock than its warrant. The assortative indexes are left-skewed, except for $e_s$ of the stock. The kurtosis of the two metrics are slightly larger than 3 for the stock and significantly larger than 3 for the warrant.

The average path lengthes of the stock and its warrant trading networks equal to $4.80 \pm 2.51$ and $6.95 \pm 1.56$, respectively. Their distributions are right-skewed and exhibit fat tails.

\section{Financial variables}
\label{Section:MarketCharacteristic}

We employ four traditional financial variables (return, volatility, intertrade duration, and trading volume) to depict the market behaviors in each windows. All the four variables contain valuable information on the underlying price formation process.

Return $r$ is defined as the difference between the logarithmic price right after the last transaction and the logarithmic price right before the first transaction in each time window. Volatility $v$ is defined as the difference between the maximum logarithmic price and the minimum logarithmic price  in each time window. Intertrade duration $\tau$ is defined as the mean value of the time elapsed between two consecutive transactions in each time window. Trading volume $w$ is defined as the total number of shares both bought and sold during each time window. By performing Augmented Dickey-Fuller tests, Jarque-Beta tests, and Ljung-Box Q-tests, we conclude that all the financial variables are stationary, do not obey normal distributions, and possess some autocorrelations, except that the return time series does not exhibit long memory.

Table~\ref{Tb:Statistics:FV} reports the summary statistics of the four financial variables. Compared with the stock, one can find that the warrant has larger volatilities, higher trading activities, and larger trading volumes. We also find that the distribution of the stock returns is almost symmetric, while the distribution of the warrant returns is left-skewed. The fact that the kurtosis of the financial variables are greater than 3 indicating that their distributions have fat tails, which are consistent with precious works on stocks and warrant for returns and volatility \citep{Gopikrishnan-Meyer-Amaral-Stanley-1998-EPJB,Gabaix-Gopikrishnan-Plerou-Stanley-2003-Nature,Gu-Chen-Zhou-2008a-PA}, intertrade \citep{Ivanov-Yuen-Podobnik-Lee-2004-PRE,Jiang-Chen-Zhou-2008-PA,Ruan-Zhou-2011-PA}, and trading volume \citep{Mu-Chen-Kertesz-Zhou-2009-EPJB}.

\setlength\tabcolsep{3.5pt}
\begin{table}[tp]
\begin{center}
  \caption{\label{Tb:Statistics:FV} Summary statistics of financial variables.}
  \medskip
\begin{tabular}{ccccc}
  \hline\hline
     & $r$ & $v$ & $\tau$ & $w$  \\
     \hline
 \multicolumn{5}{l}{Panel A: Stock} \\
 Mean & 0.0000 & 0.0038 & 6.7599 & 967173 \\
 Median & 0.0000 & 0.0025 & 5.5962 & 586600 \\
 Maximum & 0.0182 & 0.0291 & 69.750 & 19186704 \\
 Minimum & -0.0246 & 0.0000 & 0.2407 & 8350 \\
 Std. Dev. & 0.0024 & 0.0020 & 4.6957 & 1226134 \\
 Skewness & 0.1078 & 2.7148 & 1.7941 & 4.4427 \\
 Kurtosis & 8.3980 & 19.0289 & 9.8683 & 36.3647 \\
 \hline
 \multicolumn{5}{l}{Panel B: Warrant} \\
 Mean & -0.0005 & 0.0138 & 1.1990 & 11995271 \\
 Median & 0.0000 & 0.0096 & 0.9492 & 6442832 \\
 Maximum & 0.1210 & 0.9520 & 9.7333 & 180404593 \\
 Minimum & -0.7492 & 0.0000 & 0.0366 & 51411 \\
 Std. Dev. & 0.0149 & 0.0176 & 0.9749 & 14924879 \\
 Skewness & -15.020 & 19.285 & 1.5684 & 3.0154 \\
 Kurtosis & 693.20 & 854.47 & 6.7192 & 15.8798 \\
  \hline\hline
\end{tabular}
\end{center}
\end{table}

\section{Correlations between financial variables and topological metrics of trading networks}
\label{Section:Correlations}

To reveal the connection between the topological structure of trading networks and the behaviors of the stock and its warrant, we estimate the correlation coefficients between the four financial variables and the five network metrics. The results are reported in Table~\ref{Tb:Results:Correlations}.

The returns of both securities are positively correlated with the centralization measures $C_k$ and $C_s$ at the significance level of 0.1\%. A trading network with higher centralization has more dominating buyers and such kind of network structures are resulted from one or a few large market orders, which also leads to positive returns \citep{Lillo-Farmer-Mantegna-2003-Nature,Zhou-2012-QF}. Hence, the price is more likely to increase consecutively and results in a large positive return. The stock returns are positively correlated with the assortative indexes $e_k$ and $e_s$, but uncorrelated with the average path length. In contrast, the warrant returns are negatively correlated with the average path length, but uncorrelated with the assortative indexes $e_k$ and $e_s$.

\setlength\tabcolsep{3.5pt}
\begin{table}[tp]
\begin{center}
  \caption{\label{Tb:Results:Correlations} Correlations between financial variables and topological metrics of trading networks. $^{***}$ significant at 0.1\%; $^{**}$ significant at 1\%; $^{*}$ significant at 5\%.}
  \medskip
\begin{tabular}{ccccccc}
  \hline\hline
     & $C_k$ & $C_s$ & $e_k$ & $e_s$ & $l$  \\
\hline
\multicolumn{6}{l}{Panel A: Stock} \\
  $r$    & $~~~0.253^{***}$ & $~~~0.188^{***}$ & $~~~0.037^{***}$ & $~~~0.031^{***}$ & $~~0.018^{~~~}$ \\
  $v$    & $  -0.015^{~~~}$ & $  -0.012^{~~~}$ & $~~~0.302^{***}$ & $~~~0.051^{***}$ & $  -0.119^{***}$ \\
  $\tau$ & $~~~0.009^{~~~}$ & $  -0.011^{~~~}$ & $  -0.382^{***}$ & $ ~~0.008^{~~~}$ & $  -0.383^{***}$ \\
  $w$    & $~~~0.014^{~~~}$ & $~~~0.035^{***}$ & $~~~0.185^{***}$ & $  -0.064^{***}$ & $~~~0.070^{***}$ \\
\hline
\multicolumn{6}{l}{Panel B: Warrant} \\
  $r$    & $~~~0.131^{***}$ & $~~~0.056^{***}$ & $  -0.003^{~~~}$ & $  -0.017^{~~~}$ & $-0.043^{***}$ \\
  $v$    & $~~~0.016^{~~~}$ & $~~~0.003^{~~~}$ & $~~~0.261^{***}$ & $~~~0.072^{***}$ & $-0.095^{***}$ \\
  $\tau$ & $  -0.027^{**~}$ & $~~~0.007^{~~~}$ & $  -0.561^{***}$ & $  -0.049^{***}$ & $-0.046^{***}$ \\
  $w$    & $~~~0.059^{***}$ & $~~~0.009^{~~~}$ & $~~~0.379^{***}$ & $~~~0.019^{*~~}$ & $-0.183^{***}$ \\
  \hline\hline
\end{tabular}
\end{center}
\end{table}

For both securities, the volatility is positively correlated with the assortative indexes $e_k$ and $e_s$, which is in sharp contrast to the negative correlation between volatility and assortative index in E-mini S\&P 500 futures market \citep{Adamic-Brunetti-Harris-Kirilenko-2012-SSRN}, and negatively correlated with the average path length $l$. More than 99.98\% of our trading networks are disassortative, which means that the nodes with larger degrees prefer to connect with the nodes with small degrees. However, by checking the local structures of the nodes with large degrees, we surprisingly find that those nodes having large degrees are connected by multiple edges in some trading networks, known as an abnormal trading motif \citep{Jiang-Xie-Xiong-Zhang-Zhang-Zhou-2013-QFL}. Such trading networks may exhibit relative large assortative indexes, but still negative. Obviously, the abnormal trading activities will increase the market volatilities and adding bridges between hubs will absolutely shorten the average path length in trading networks, which also leads to negative correlations between the market volatility and the average path length in trading networks.

The stock intertrade duration $\tau$ is negatively correlated with the assortative index $e_k$ and the average path length $l$, while the warrant intertrade duration is negatively correlated with the centralization $C_k$, the assortative indexes $e_k$ and $e_s$, and the average path length $l$. Shorter intertrade durations correspond to higher trading activities, which usually happen when dominating buyers and seller are lack. This could increase the assortative index of trading network. In this situation, the network usually contains a relatively large number of nodes and is less likely to have a star-like structure such that the average path length is more probable to be large.

The trading volume $w$ is positively correlated with the centralization $C_s$ for the stock and with $C_k$ for the warrant. A possible interpretation is the following. Although there is no asymmetry in the volume-return relationship at the transaction level \citep{Zhou-2012-QF}, it is well established that it takes larger trading volume to move the price up than to hit it down with the same magnitude of return when the time scale is aggregated (say 5 min as in the present work) \citep{Karpoff-1987-JFQA}. Since a high centralization corresponds to a high return, the trading volume will also be high. The correlations between trading volume and assortative index $e_k$ are positive for both securities. This is because that higher trading volume usually corresponds to large trading activities and the lack of dominating buyers and sellers, which implies higher assortative index as for small inter-trading durations. The correlations between the trading volume and the assortative index are mixed. Moreover, the correlation between the trading volume $w$ and the average path length $l$ is positive for the stock and negative for the warrant. It is hard to give an intuitive explanation for these mixed correlations.

\section{Summary}
\label{Section:Conclusion}

In this work, we investigate the statistical properties of five topological metrics of trading networks and four financial variables in the five-minute moving windows of the stock Baosteel and its warrant. We find that the Chinese security trading networks have disassortative mixing patterns and similar order execution patterns in both sides of the limit order book. The investigated financial variables are stationary and possess fat tails. We also find that, for both stock and warrant, the return, volatility and trading volume positively correlate with the centralization or assortativity, while the intertrade duration is negatively correlated with the network centralization measures and assortative indexes. Our results show that the stock and its warrant share many common topological properties in their trading networks, but also exhibit idiosyncratic characteristics.

Our findings show that topological metrics of trading networks can be used to characterize the temporal properties of financial variables and have the potential to gain a deep understanding of the price formation process from a novel perspective. It is also possible to design possible techniques to identify outliers in the time series of trading networks, which is able to underpin clues for abnormal behaviors in the price dynamics and provide information about possible price manipulations.

\bigskip
{\textbf{Acknowledgments}}
This work was partially supported by the National Natural Science Foundation of China (11075054, 71101052 and 71131007), Shanghai ``Chen Guang'' Project (2012CG34), Shanghai Rising Star (Follow-up) Program (11QH1400800), Program for Changjiang Scholars and Innovative Research Team in University (IRT1028), and the Fundamental Research Funds for the Central Universities.

\bigskip
\bibliographystyle{elsarticle-harv}

\bibliography{/home/zqjiang/research/Papers/Auxiliary/Bibliography_FullJournal}
\end{document}